\documentclass[preprint,12pt]{elsarticle}

\usepackage{amssymb}
\usepackage{xcolor}
\usepackage{graphicx}
\usepackage{subcaption}
\usepackage{hyperref}
\usepackage{amsmath}
\usepackage{array}
\usepackage{makecell}
\usepackage{geometry}
\usepackage{tabularx}
\usepackage{colortbl}
\usepackage{adjustbox}
\usepackage{longtable}
\usepackage{lscape}
\usepackage{comment}
\definecolor{darkgreen}{rgb}{0.0, 0.5, 0.0}

\journal{}
\begin{document}
\begin{frontmatter}

\title{PhishKey: A Novel Centroid-Based Approach for Enhanced Phishing Detection Using Adaptive HTML Component Extraction}

\author[label1,label2]{Felipe Castaño}\ead{lcastl02@estudiantes.unileon.es}
\author[label2,label3]{Eduardo Fidalgo}\ead{eduardo.fidalgo@unileon.es}
\author[label2,label3]{Enrique Alegre}\ead{enrique.alegre@unileon.es}
\author[label2,label3]{Rocio Alaiz-Rodríguez}\ead{rocio.alaiz@unileon.es}
\author[label1]{Raúl Orduna}\ead{rorduna@vicomtech.org}
\author[label1]{Francesco Zola}\ead{fzola@vicomtech.org}

\affiliation[label1]{organization={Vicomtech Foundation, Basque Research and Technology Alliance (BRTA)},
            city={San Sebastian},
            country={Spain}}
\affiliation[label2]{organization={Department of Electrical Engineering, Systems and Automation, Universidad de León},
            city={León},
            country={Spain}}

\affiliation[label3]{organization={Spanish National Institute of Cybersecurity (INCIBE)},
            city={León},
            country={Spain}}

\begin{abstract}
Phishing attacks pose a significant cybersecurity threat, evolving rapidly to bypass detection mechanisms and exploit human vulnerabilities. This paper introduces PhishKey to address the challenges of adaptability, robustness, and efficiency. PhishKey is a novel phishing detection method using automatic feature extraction from hybrid sources. PhishKey combines character-level processing with Convolutional Neural Networks (CNN) for URL classification, and a Centroid-Based Key Component Phishing Extractor (CAPE) for HTML content at the word level. CAPE reduces noise and ensures complete sample processing avoiding crop operations on the input data. The predictions from both modules are integrated using a soft-voting ensemble to achieve more accurate and reliable classifications.
Experimental evaluations on four state-of-the-art datasets demonstrate the effectiveness of PhishKey. It achieves up to $98.70\%$ F1 Score and shows strong resistance to adversarial manipulations such as injection attacks with minimal performance degradation. 
\end{abstract}


\end{frontmatter}

\section{Introduction}
\label{sec:intro}

Phishing is one of the most persistent and impactful enablers of cybercrime today since it is a primary social engineering method for criminals to infiltrate systems, steal sensitive information, and commit fraud. This attack primely exploits human vulnerabilities, often more difficult to protect than technical weaknesses \cite{hong2012state}. Besides, phishing attacks are being deployed on a large scale with minimal effort, as shown in the Phishing Activity Trends Report of the 3rd Quarter 2024 \cite{Anti-PhishingWorkingGroup2024}. The report highlights that the members of the Anti-Phishing Working Group (APWG)\footnote{https://apwg.org/} have reported over $300,000$ phishing attacks per month from July to September 2023. This trend is based on the application of new paradigms as well, such as Phishing-as-a-service that provides products, services, and data of victims, allowing non-expert persons to set up and launch their phishing campaigns \cite{europol2024iocta}. In this context, automated tools allow attackers to send millions of emails, texts, or messages, reaching potential victims and maximizing the chances of successful exploitation.

Phishing has evolved beyond simple emails, and new vector attacks have been identified, such as spear-phishing (targeted attacks), vishing (voice phishing), smishing (SMS phishing), quishing (QR code phishing), and social media phishing \cite{edwards2024multi}. Cybercriminals can trick victims using emerging technology, such as deepfake, through voice and video manipulation and large language models (LLMs) crafting phishing content \cite{falade2023decoding, yao2024survey}. These advances make phishing more sophisticated, convincing, and challenging to detect.

Phishing detection algorithms present different approaches and complexity, aiming to identify and mitigate phishing attacks. These algorithms can be differentiated according to the information they analyze. For example, URL-based algorithms \cite{ sahingoz2019machine, sahingoz2024dephides} focus on identifying suspicious patterns or features within URLs, such as uncommon domains, lengthy URL paths, misspelled keywords, or domain obfuscation techniques; HTML-based algorithms \cite{opara2020htmlphish, mensah2015ajna, shukla2024http}, parse the source code to look for characteristics common to phishing sites, such as the presence of form fields that ask for sensitive data (e.g., usernames and passwords), JavaScript obfuscation, or the absence of security certificates hybrid algorithms integrate features from both URL and HTML contexts to enhance their ability to detect potential phishing \cite{ opara2024look, prasad2024phiusiil, sanchez2022phishing}; finally, other detection algorithms rely on computer vision and image processing techniques to compare the screenshot of a suspect webpage to known legitimate pages or phishing templates.

In the same way, the detection algorithms can be separated according to how the features are extracted and analyzed \cite{asif2023machine, asiri2023survey}. Among them, we can find handcraft algorithms using manually extracted features \cite{sahingoz2019machine, sanchez2022phishing}, following guidelines and approaches identified by security experts. Other approaches focus on automatically extracting features using machine learning and deep learning techniques \cite{bozkir2023grambeddings, opara2024look}. Additionally, hybrid approaches \cite{rao2020catchphish, sahingoz2024dephides} combine the strengths of two previously mentioned strategies.

Despite the progress in phishing detection, several challenges and unresolved issues remain. First, many methods lose essential information during preprocessing, such as HTML inputs shortened or cropped \cite{ariyadasa2022combining, opara2024look}. Additionally, phishing attacks constantly evolve, with attackers frequently changing their tactics to bypass detection systems. Finally, existing methods are characterized by low robustness in the presence of adversarial manipulations—such as data injection or subtle alterations—which can significantly degrade the model detection performance. 
To tackle these challenges, we present PhishKey, a novel deep-learning framework for phishing detection inspired by the text summary method proposed by Rossielo et al. \cite{Rossiello2017}, where the more representative words in a sample are extracted using a centroid extractive technique. 

The contributions of this paper can be summarized as follows:

\begin{itemize}
    \item We propose PhishKey, a deep-learning framework for phishing detection combining character-level processing of URLs and word-level analysis of HTML sources,  relying on automatic feature extraction to adapt to evolving attacks. This method is inspired by Opara et al. \cite{opara2024look}, enhancing the robustness against adversarial manipulations, keeping the full sample content, and combining predictions with a soft-voting method.
    \item We propose CAPE (Centroid-Based Key Component Phishing Extractor), a clustering-based approach that leverages word embeddings and centroid extractive summarization to enhance phishing detection accuracy and efficiency, avoiding lost information due to padding and cropping operations. Inspired by the method proposed by Rossiello et al. \cite{Rossiello2017}, CAPE innovatively adapts centroid-based techniques to the HTML domain of phishing detection by grouping representative words, calculating class-specific centroids, and extracting key components from HTML samples.
    \item Finally, we evaluated our method using four state-of-the-art datasets due to the lack of a common dataset that can be used as a reference. This evaluation assesses the ability of the model to handle diverse legitimate and phishing samples and compares PhishKey results in several test scenarios with methods using an automated extraction approach. 
\end{itemize}

The remaining sections of this paper are organized as follows: Section \ref{sec:background} provides a comprehensive review of the state of the art. In Section \ref{sec:key_component}, we introduce our method, called key component extraction, by describing its design and implementation. Section \ref{sec:experimental} explains the experimental setup, including the dataset and evaluation metrics. The results of our experiments are analyzed and discussed in Section \ref{sec:results}, where we evaluate the effectiveness of the proposed method. Finally, Section \ref{sec:Conclusions} summarizes key findings and suggests possible directions for future research.

\section{Literature review}
\label{sec:background}
This section reviews recent and relevant methods for detecting phishing attacks on websites. Due to the ever-evolving tactics of cybercriminals, this is a very dynamic and challenging area of research. 

We consider the input and its preprocessing crucial in a constantly evolving environment. Consequently, we have categorized phishing attack methods in terms of how authors extracted features: First, handcrafted features rely on manually extracted characteristics, utilizing the knowledge and expertise of security researchers to identify relevant patterns. Then, automatic feature extraction employs artificial intelligence methods, such as embeddings, to extract features from raw data without manual intervention. Finally, hybrid approaches aim to combine the strengths of both manual and automated methods.
  
\subsection{Handcraft feature extraction}
Several studies have employed handcrafted features for phishing detection. In \cite{aljofey2022effective}, the authors propose extracting novel features from web pages based on the URL character sequence, textual content (using the Term Frequency-Inverse Document Frequency, TF-IDF algorithm), and hyperlinks to determine the relationship between the content and the URL of a web page. After feature extraction,  the authors evaluate several classification algorithms such as Random Forest (RF), Decision Tree (DT), LightGBM  (LGBM), Logistic Regression (LR), Support Vector Machine (SVM), and XGBoost (XGB). The study used a dataset containing $32,972$ benign and $27,280$ phishing web pages. The results showed that this approach achieved an accuracy of $96.76\%$, outperforming some baseline approaches. 

Similarly, Bahnsen et al. \cite{bahnsen2017classifying} used $14$ handcrafted features extracted from URLs for phishing attack detection. They trained and tested different models, including RF and Long-Short Term Memory (LSTM) networks, incorporating some feature selection mechanisms. The LSTM model achieved the best performance, with an accuracy of $98.70\%$. However, this approach required more time for training and a more complex process for tuning the network parameters.

Karim et al. \cite{karim2023phishing} used handcrafted features to train a hybrid model composed of Linear Regression, Support Vector Machine, and Decision Tree classifiers, using both soft and hard voting mechanisms. The study showed that this new model reduces the execution time and increases the accuracy in defending against phishing attacks. However, this hybrid model is not the best in F1-score among the approaches evaluated in the paper.

In \cite{niakanlahiji2018phishmon}, the authors introduce PhishMon, a Random Forest model that distinguishes between legitimate and phishing web pages. The tool is trained with 15 handcrafted features extracted from web pages and URLs without requiring third-party services. The authors state that the proposed 15 features are costly for phishers to replicate, making them reliable for classification. PhishMon has been validated using a dataset containing 4,800 phishing and 17,500 legitimate web pages, showing an accuracy of $95.40\%$.

A similar approach is proposed in \cite{li2019stacking}, where the authors combine URL and HTML features - also excluding third-party service data -  and stack three machine learning (ML) models in multiple layers: Gradient Boosting Decision Tree (GBDT), XGB, and LGBM. This operation allows them to have complementary models for improving the classification task, and the authors reported an accuracy of 97.30\%.

An approach that combines URL clustering, classification, and categorization for ranking URL samples is introduced in \cite{feroz2015phishing}. Specifically, the work uses lexical and host-based features, i.e., metadata extracted from analyzing the URL using third-party services such as IP, DNS, and NS. The approach was validated using samples collected from DMOZ\footnote{\url{https://www.dmoz.co.uk/}} and PhishTank\footnote{\url{https://phishtank.org/}}. 

In \cite{prasad2024phiusiil}, the authors proposed a phishing detection model based on URL feature analysis. These handcrafted features were used to compute similarity indexes and implement incremental learning. According to the authors, the former should help detect attacks such as zero-width characters \cite{bashir2020high}, homographs \cite{holgers2006cutting}, and Punycode \cite{abawajy2018securing}, while the latter should allow the framework to be updated continuously. The authors reported that the approach reached an accuracy of about $99.24\%$ using a dataset containing 100945  phishing samples and 134850 legitimate samples.

Sahingoz et al. \cite{sahingoz2019machine} presented a real-time system for detecting phishing websites through URL analysis. The study evaluated seven classification algorithms with natural language processing (NLP)-based features. The authors built a dataset to assess the methods. Phishing URL samples were collected from PhishTank, while legitimate URLs were collected from Yandex Search API \footnote{\url{https://yandex.com.tr/dev/xml}}. The dataset contains 37,175 phishing and 36,400 legitimate samples. Among the algorithms evaluated  - NB, RF, kNN, ADA, K-star, Sequential Minimal Optimization (SMO), and DT- Radom Forest (RF) achieved the best performance, with an accuracy of 97.98\%.

A more recent study by Sanchez-Paniagua et al. \cite{sanchez2022phishing} addressed a critical issue with existing dataset collection methods, highlighting that many rely on outdated datasets created from the homepages instead of the login webpages and fail to identify phishing websites that mimic legitimate login pages effectively. The authors proposed 54 features, grouped into four categories: URL, HTML, hybrid, and technological features. They evaluated several machine learning models, such as SVM, LR, NB, RF, kNN, and ADA, and two powerful ensemble methods, LGBM and XGB. To mitigate the exposed problem, the authors compiled the Phishing Index Login Websites Dataset (PILWD-134K) containing 134,000 verified phishing and legitimate website samples. Among the models assessed, LGBM achieved the highest accuracy at $97.95\%$.

\subsection{Automatic Feature Extraction}
\label{sub:automatic}

In 2020, Opara et al. \cite{opara2020htmlphish} introduced HTMLPhish, an approach for classifying phishing websites that focus on HTML content without extensive handcraft feature engineering. In the preprocessing step, the samples were tokenized into individual characters or words, including punctuation, as separate tokens; finally, the sequences were padded to a fixed length of 180 characters. The approach uses a CNN to capture semantic dependencies within the HTML content. The authors collected two sets of data: D1, with 23,000 legitimate and 2,300 phishing pages, crawled in November 2018, and D2, with 24,000 legitimate and 2,400 phishing pages, crawled in January 2019. The results showed the model achieved $98.00\%$ accuracy on D1 and $93.00\%$ on D2, demonstrating robustness to temporal changes in phishing techniques. 

In \cite{ariyadasa2022combining}, the authors use the Long-term Recurrent Convolutional Network (LRCN) for URL analysis and the Graph Convolutional Network (GCN) for HTML content analysis. The study introduces a representation to automatically analyze both the URL and HTML content of a webpage; on the one hand, the URL is tokenized at the character level and normalized to a maximum length of $150$ characters, with longer ones truncated and shorter ones padded. On the other hand, HTML content is transformed into a graph where nodes represent HTML tags and attributes. Finally, the results obtained by LRCN and GCN are combined to get the final classification. The approach was evaluated using three public datasets: Dataset A ($20,000$ phishing and $20,000$ legitimate sites), Dataset B ($25,000$ phishing and $25,000$ legitimate sites), and a Benchmark Dataset ($24,800$ legitimate and 21,296 phishing sites). The authors reported an accuracy of $96.42\%$ on Dataset B.

Bozkir et al. \cite{bozkir2023grambeddings} explored an automated n-gram selection and a filtering mechanism with URL feature extraction. Their approach combines the features using a cascading approach that concatenates CNN, LSTM, and Attention layers. This method outperformed other literature models, showing an accuracy of $98.27\%$ on a dataset collected by the authors to evaluate the model. The dataset includes Phishing URLs collected between May 2019 and June 2021, resulting in a balanced dataset with $400,000$ phishing URLs and $400,000$ legitimate URLs. 

Opara et al. \cite{opara2024look} later improved their previous method by presenting WebPhish, which incorporates a hybrid input methodology, including the URL. The model uses the raw data at a character level with sequences padded or truncated to fixed lengths of 180 characters for URL and 2000 words for HTML. The authors used a Deep Neural Network (DNN) architecture, including embedding layers, concatenation, Convolutional Neural Network  (CNN), and fully connected layers. The authors collected a dataset of $22,687$ legitimate and $22,687$ phishing pages to evaluate the results, reporting an accuracy of $98.10\%$. 

\subsection{Hybrid Feature Extraction}

In \cite{rao2020catchphish}, the authors assessed the performance of several ML models - XGB, RF, LR, kNN, SVM, and DT- to predict the legitimacy of web pages by only investigating the URL information. In particular, they compared three cases: (1) using handcrafted URL features, (2)  using TF-IDF features, and (3) combining both feature sets. They conclude that using the third approach; they can outperform the results obtained in two previous works \cite{sahingoz2019machine, marchal2014phishstorm}.

Additionally, \cite{korkmaz2022hybrid} introduced a two-step approach for phishing detection. In the first step, a Generative Convolutional Neural Network (GCNN) classifies handcrafted and character embedding features extracted from URLs. In the second step, a DNN model with content-based handcrafted features is used.  

Hybrid feature extraction methods combine different approaches to improve the performance of phishing detection systems by integrating handcrafted and automated techniques.
The work by \cite{sahingoz2024dephides} compares five deep learning algorithms: Multi-Layer Perceptron (MLP), CNN, Recurrent Neural Networks (RNN), Bidirectional Recurrent Neural Network (BRNN), and Attention Networks (ATT) to detect phishing web pages by analyzing their URLs. In particular, each URL was encapsulated in 200 characters using truncation or padding operations. A character-based embedding approach was employed for the feature extraction process (vectorization). The study was validated on a dataset of 5.1 million entries, with 2.3 million instances gathered from PhishTank (phishing) and the remaining from Common Crawl\footnote{https://commoncrawl.org/} (legitimate). The results showed that the CNN algorithm achieved an F1-score of $98.74\%$, outperforming traditional machine learning algorithms.

This comparative overview, as shown in Table \ref{tab:summarization}, allows for a clearer understanding of the approaches and highlights the lack of a common dataset that can be used as a reference, as most authors collect their data independently. This variation in datasets makes it challenging to compare the approaches, as each author evaluated their method using a different data set.

\begin{table}[t]
\centering
\begin{adjustbox}{max width=\linewidth}
\begin{tabular}{|>{\centering\arraybackslash}m{2.5cm}|>{\centering\arraybackslash}m{2.5cm}|>{\centering\arraybackslash}m{2.5cm}|>{\centering\arraybackslash}m{4cm}|>{\centering\arraybackslash}m{2cm}|>{\centering\arraybackslash}m{2.3cm}|>{\centering\arraybackslash}m{2.3cm}|>{\centering\arraybackslash}m{1.5cm}|>{\centering\arraybackslash}m{2.3cm}|>{\centering\arraybackslash}m{1.3cm}|>{\centering\arraybackslash}m{1.3cm}|}
\hline
\textbf{Bibtex} & \textbf{Extraction} & \textbf{Input} & \textbf{Algorithm} & \textbf{Dataset Source} & \textbf{Legitimate Samples} & \textbf{Phishing Samples} & \textbf{Public} & \textbf{Collection Time} & \textbf{ACC} & \textbf{F1} \\ \hline
\textcolor{red}{Aljofey et al. \cite{aljofey2022effective}} & \textcolor{red}{Handcraft} & \textcolor{red}{Hybrid Inp.} & \textcolor{red}{RF, DT, LGBM, LR, SVM, XGB} & \textcolor{red}{Collected} & \textcolor{red}{32972} & \textcolor{red}{27280} & \textcolor{red}{Yes} & \textcolor{red}{2016} & \textcolor{red}{96.76} & \textcolor{red}{96.38} \\ \hline
Bahnsen et al. \cite{bahnsen2017classifying} & Handcraft & URL & RF, LSTM & Collected & 1M & 1M & No & No reported & 98.7 & 98.76 \\ \hline
Karim et al. \cite{karim2023phishing} & Handcraft & URL & LR, SVM, DT & Collected & 4500 & 6500 & Yes & 2022 & 95.41 & 95.91 \\ \hline
Niakanlahiji et al. \cite{niakanlahiji2018phishmon} & Handcraft & Hybrid Inp. & CART, KNN, AdaBoost, RF & Collected & 4807 & 17508 & No & 2017 & 95.4 & 88.68 \\ \hline
\textcolor{red}{Li et al. \cite{li2019stacking}} & \textcolor{red}{Handcraft} & \textcolor{red}{Hybrid Inp.} & \textcolor{red}{GBDT, XGB, LGBM} & \textcolor{red}{Collected} & \textcolor{red}{30873} & \textcolor{red}{19074} & \textcolor{red}{Yes} & \textcolor{red}{2009-2017} & \textcolor{red}{97.3} & \textcolor{red}{No reported} \\ \hline
\textcolor{red}{Prasad et al. \cite{prasad2024phiusiil}} & \textcolor{red}{Handcraft} & \textcolor{red}{Hybrid Inp.} & \textcolor{red}{RF, DT, LGBM, XGB} & \textcolor{red}{Collected} & \textcolor{red}{134850} & \textcolor{red}{100945} & \textcolor{red}{Yes} & \textcolor{red}{2022-2023} & \textcolor{red}{99.24} & \textcolor{red}{No reported} \\ \hline
Sahingoz et al. \cite{sahingoz2019machine} & Handcraft & URL & NB, RF, kNN, Adaboost, K-star, SMO, DT & Collected & 36400 & 37175 & Yes & 2017 & 97.98 & 98 \\ \hline
\textcolor{red}{Sanchez-Paniagua et al. \cite{sanchez2022phishing}} & \textcolor{red}{Handcraft} & \textcolor{red}{Hybrid Inp.} & \textcolor{red}{SVM, LR, NB, RF, kNN, ADA} & \textcolor{red}{Collected} & \textcolor{red}{66000} & \textcolor{red}{66000} & \textcolor{red}{Yes} & \textcolor{red}{2020} & \textcolor{red}{97.95} & \textcolor{red}{98} \\ \hline
\rowcolor{gray!20} Bozkir et al. \cite{bozkir2023grambeddings} & Automated & URL & CNN, LSTM, ATT & Collected & 400000 & 400000 & Yes & 2021 & 98.27 & 98.26 \\ \hline
\rowcolor{gray!20} Opara et al. \cite{opara2020htmlphish} & Automated & HTML & CNN & Collected & 50000 & 4700 & No & 2019 & 98 & 97 \\ \hline
\rowcolor{gray!20} \textcolor{red}{Opara et al. \cite{opara2024look}} & \textcolor{red}{Automated} & \textcolor{red}{Hybrid Inp.} & \textcolor{red}{CNN} & \textcolor{red}{Collected} & \textcolor{red}{22687} & \textcolor{red}{22687} & \textcolor{red}{Yes} & \textcolor{red}{2020} & \textcolor{red}{98.1} & \textcolor{red}{98.1} \\ \hline
\rowcolor{gray!20} Ariyadasa et al. \cite{ariyadasa2022combining} & Automated & Hybrid Inp. & LRCN, GCN & Public & 25000 & 25000 & Yes & 2019 & 96.42 & 96.42 \\ \hline
Sahingoz et al. \cite{sahingoz2024dephides} & Hybrid Extraction & URL & MLP, CNN, RNN, BRNN, ATT & Collected & 2881948 & 2320893 & Yes & 2018 & 98.74 & 98.74 \\ \hline
Rao et al. \cite{rao2020catchphish} & Hybrid Extraction & URL & XGB, RF, LR, KNN, SVM, DT & Collected & 85409 & 40668 & Yes & No reported & 94.26 & 95.88 \\ \hline
Kormaz et al. \cite{korkmaz2022hybrid} & Hybrid Extraction & URL & GCNN, DNN & Collected & 51316 & 36173 & No & 2006-2021 & 98.37 & No reported \\ \hline
\end{tabular}
\end{adjustbox}
\caption{Summary table of the state-of-the-art approaches for phishing classification. Highlighted in red are the methods that used hybrid inputs and have publicly available datasets. On gray are the methods that use an automated feature extraction}
\label{tab:summarization}
\end{table}

However, certain studies, like those by Li et al. \cite{li2019stacking} and Prasad et al. \cite{prasad2024phiusiil}, only shared their data after preprocessing, providing only the extracted features, which makes these datasets unsuitable for this study. In contrast, datasets from Cui et al. \cite{cui2017tracking}, Sanchez-Paniagua \cite{sanchez2022phishing}, and Opara et al. \cite{opara2024look} are incorporated into the analysis and evaluation of the proposed method.

\section{PhishKey Method}
\label{sec:key_component}

Phishing attacks evolve continuously to avoid detection, outdating models quickly and making it necessary to update them \cite{sanchez2021impact}. From this perspective, handcrafted feature models are often ad hoc solutions, relying heavily on the ability of the researcher to identify relevant features, and retraining such models is challenging. In contrast, automatic feature extraction methods can adapt to changes in the phishing attack landscape since the network receives the raw data directly, training the model to identify the relevant parts of the input by itself. This approach facilitates retraining models with minimal changes as new samples emerge. For those reasons, we decided to focus on automatic feature extraction, which could enhance the generalization of the proposed models and the adaptation to new threats without significant modifications.

We aim to address the challenges in existing methods, particularly the loss of information caused by preprocessing steps such as cropping, as seen in the approach by Opara et al. \cite{opara2024look} where the HTML input is cropped to the first 2000 words. While URL content is short and does not vary much regarding the number of characters used, HTML content can be more complex, and cropping can lead to the loss of important information. To address this, we propose CAPE (Centroid-Based Key Component Phishing Extractor), a clustering-based approach that leverages word embeddings and centroid extractive summarization to enhance phishing detection reducing lost information due to padding and cropping operations. CAPE is at the core of PhishKey and is the main contribution of this work, and it is explained in detail in Section \ref{sec:cape}.

PhishKey method has two separate classification modules: one for classifying URLs and the other for classifying the HTML source of a webpage. Each module processes its data independently, providing insights into potential phishing sites. The final decision is made by combining the predictions from both modules in an ensemble method, ensuring a more reliable result.
 
As mentioned earlier, current models using automated feature extraction from URLs can efficiently process the entire sample. Due to their limited corpus and -length, models can process the whole URL with minimal variation. Therefore, we decided to implement the approach proposed by Opara et al. \cite{opara2024look} in the URL-based sample classification step. This implementation processes the URL at the character level, using an embedding layer before the classification layers of the model, allowing the model to receive the complete samples without previous steps. Specifically, the model for URL classification includes the following layers: An embedding layer using a dimension of 16, in charge of transforming the character of the raw URL samples into a feature vector. Next, the second component is a 1D convolutional layer that uses the results of the first component as input. Finally, the last component is a fully connected layer that receives the information from the CNN and max-pooling layers for a classification result.

\subsection{CAPE: Centroid-Based Key Component Phishing Extractor}
\label{sec:cape}

We developed a custom tokenizer to handle the difficulties of HTML structures and embedded code. The tokenizer is built to parse the HTML samples by handling various elements typically contained in the web page content, such as tags, attributes, JavaScript, and CSS. The tokenizer performs preprocessing steps to normalize and clean the HTML data. In parallel, we address structural patterns within the HTML to isolate each tag or element and guarantee they are processed as a distinct element during the tokenization phase. Finally, the code is split into tokens.

In addition to processing the main HTML structure, the tokenizer handles JavaScript and CSS code, isolating them by following the previous process but adapted for non-HTML content. This extra step ensures that the code embedded within the HTML file is tokenized correctly and included in the analysis. 

Once the tokens are extracted, the next step is to transform the words into embeddings, a numerical representation that allows us to perform operations on the data. An embedding maps tokens to dense vectors in a continuous vector space; these embeddings capture semantic relationships between words, allowing the model to understand the individual words and the context. We use the Word2Vec algorithm \cite{mikolov2013efficient} to generate word embeddings through unsupervised learning. This is a continuous vector of $100$ elements for each word. 

The next step in the classification pipeline is the key component extraction technique. These techniques mitigate data loss and can process the entire sample. Several state-of-the-art algorithms are used for these tasks such as BERT \cite{bharathi2023text}, GPT \cite{basyal2023text}, Pointer-Generator Networks \cite{see2017get}, and Bidirectional and Auto-Regressive Transformers (BART) \cite{xu2022sequence}. However, these models are typically pre-trained on natural language data and require considerable data to adapt to other types of content. As a result, they face limitations when applied to scenarios outside natural language contexts. For instance, HTML is more than simple text; it includes tags, CSS, and JavaScript, which makes it challenging for these models to process effectively. For this reason, we decided to utilize a more adaptable and straightforward technique for feature extraction, such as TF-IDF \cite{christian2016single, rani2021comparative} and clustering \cite{khan2019extractive}.

We implemented a clustering-based approach to group representative words, calculate the centroids of each class, and use these centroids to select the most relevant parts from each sample. This approach applies the insights from the method proposed by Rossiello et al. \cite{Rossiello2017}, who employed a technique based on centroid extractive summarization to identify and extract the more representative words. Finally, they score the relevance of the sentences according to the representative words they contain. In our implementation, we set a centroid for the class using the 100-dimension embedding representation for each word, identifying the most representative words and focusing on the most informative features by class. This method reduces noises, normalizes the input length without losing information, and guarantees that we capture the essence of each class without depending on the entire HTML sample.

After calculating the centroids for each class, the samples are processed individually. We select 2000 words from these embeddings based on their proximity to the legitimate and phishing centroids. Finally, a Bag of Words (BoW) representation is created from the selected words. This step normalizes the input while retaining the important features necessary for phishing detection.

\subsubsection{Obtaining Embeddings}
Let $D = \{d_1, d_2, \dots, d_N\}$ be the dataset consisting of HTML samples, where each $d_i$ represents a webpage (either legitimate or phishing). We use Word2Vec embedding model to represent each token $t_{i,j}$ in $d_i$ as a $k$-dimensional vector:
\begin{equation}
\mathbf{v}_{i,j} = f_{\text{emb}}(t_{i,j}), \quad \mathbf{v}_{i,j} \in \mathbb{R}^k,
\label{eq:embeddings}
\end{equation}
where $f_{\text{emb}}$ is the embedding generation function.

\subsubsection{Clustering and Centroid Calculation}
Each sample $d_i$ belongs to a class $c \in \{C_{\text{phishing}}, C_{\text{legitimate}}\}$. The token embeddings $\mathbf{v}_{i,j}$ are grouped using $k$-means to form representative clusters for phishing and legitimate websites. The centroid for each class $c$ is calculated as:
\begin{equation}
\mathbf{c}_c = \frac{1}{|T_c|} \sum_{\mathbf{v}_{i,j} \in T_c} \mathbf{v}_{i,j},
\label{eq:centroid}
\end{equation}
where $T_c$ is the set of embeddings associated with class $c$.

\subsubsection{Key Component Selection}
For each sample $d_i$, the $m$ most representative tokens are selected based on their proximity to the centroid of their respective class. The proximity is measured using the cosine similarity metric (CSM):
\begin{equation}
\text{sim}(\mathbf{v}_{i,j}, \mathbf{c}_c) = \frac{\mathbf{v}_{i,j} \cdot \mathbf{c}_c}{\|\mathbf{v}_{i,j}\| \|\mathbf{c}_c\|}.
\label{eq:similarity}
\end{equation}
The set of selected tokens is defined as:
\begin{equation}
T_{\text{selected}} = \text{top-}m\{\text{sim}(\mathbf{v}_{i,j}, \mathbf{c}_c) \mid \mathbf{v}_{i,j} \in d_i\}.
\label{eq:selection}
\end{equation}

\subsubsection{Bag of Words Representation}
Using the selected tokens, we construct a Bag of Words (BoW) representation to normalize the input length. Let $B = \{b_1, b_2, \dots, b_M\}$ be the vocabulary generated from the $m$ most representative tokens, then the BoW representation of $d_i$ is a vector:
\begin{equation}
\mathbf{b}_i = [f_{b_1}(d_i), f_{b_2}(d_i), \dots, f_{b_M}(d_i)],
\label{eq:bow}
\end{equation}
where $f_{b_k}(d_i)$ is the frequency of token $b_k$ in $d_i$.

\subsubsection{Sample Classification}
Finally, the BoW representations are used as input for a supervised classifier, the best classifier for this task will be selected on the first experiment described in Section \ref{sec:experimental}. The prediction for a sample $d_i$ is given by:
\begin{equation}
\hat{y}_i = f_{\text{clf}}(\mathbf{b}_i),
\label{eq:classification}
\end{equation}
where $f_{\text{clf}}$ is the classification function.

\subsection{Ensemble Model}
\begin{figure}[ht]
\centerline{\includegraphics[width=0.9\linewidth]{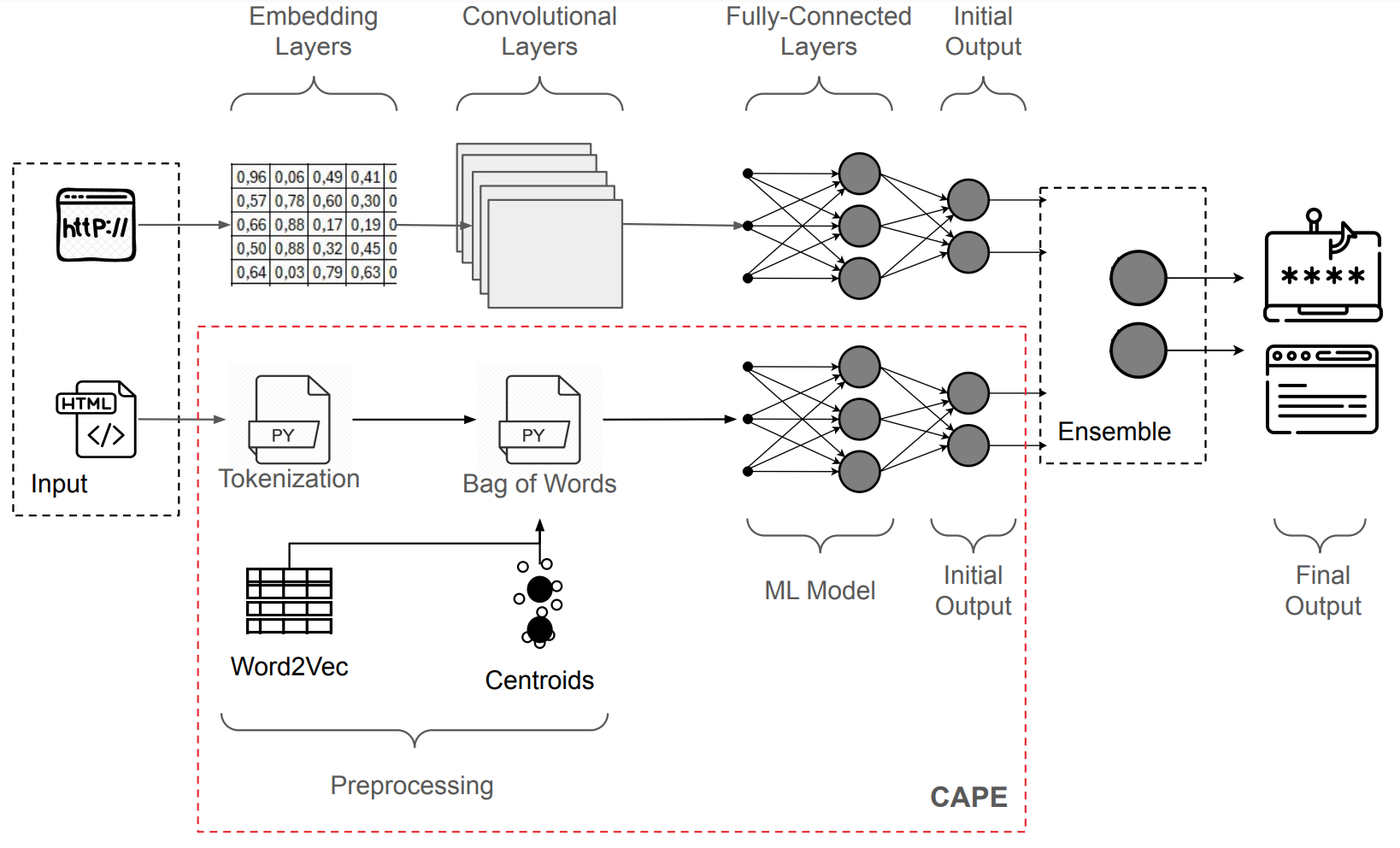}}
\caption{PhishKey components interaction. The red box contain the details of CAPE.}
\label{fig:COPE}
\end{figure}

In the final step of our pipeline, we propose an ensemble voting approach to improve prediction accuracy and robustness by integrating the predictions from the CNN model for the URL input and the Key Component Extractor model for HTML.

The ensemble model applies a soft-voting method, where the probability outputs of each model are weighted and summed to determine the final prediction. This approach allows each model to contribute proportionally based on its performance. We optimize the voting weights by conducting a grid search on validation data and adjusting the influence of each component to maximize the performance.

This ensemble approach uses URL and HTML inputs and processes each sample fully without cropping data, enhancing the robustness of the model. This final ensemble prediction model forms the complete method called PhishKey, which is designed to deliver a classification for webpage phishing detection.

\section{Experimental Setup}
\label{sec:experimental}
\subsection{Datasets}
We selected four state-of-the-art datasets containing URL and HTML resources to evaluate our proposed hybrid methodology. These datasets comprehensively assess the ability of the models to handle diverse phishing and legitimate samples. Each dataset offers a distinctive collection of data, varying in size and time of collection as shown in Table \ref{tab:phishing-dataset}. 

Cui et al. \cite{cui2017tracking} created a dataset of phishing attacks from the PhishTank platform. This dataset, collected in 2016, includes 19,066 verified phishing sites and 24,800 legitimate samples gathered from Alexa top-ranked sites. It is worth highlighting that it is one of the first datasets to release raw HTML samples without preprocessing, allowing for direct comparison between different HTML processing methods.

The PILWD dataset, proposed by Sanchez-Paniagua et al. \cite{sanchez2022phishing}, was collected between 2019 and 2020; it includes 132,000 samples, divided between phishing and legitimate data, with around 66,000 samples in each category. They used Quantcast Top Sites and The Majestic Million to identify domains for legitimate web pages, assuming that phishing sites do not appear in these lists. Finally, in the phishing web page collection step, the authors used Phishtank as the source of URLs. 

In 2019, Opara et al. \cite{opara2024look} created a dataset of 45,374 samples, evenly split between phishing and legitimate sites. The legitimate samples were obtained from Alexa top-ranked sites, while phishing samples were sourced from Phishtank..

Finally, Aljofey et al. \cite{aljofey2022effective} collected a dataset in 2020 with 60,252 samples ( 27,280 phishing and 32,972 legitimate web pages). Legitimate samples were collected from Stuf Gate42 in 2020, while the phishing samples were retrieved from PhishTank between August 2016 and April 2020. 

\begin{table}[htbp]
\centering
\rowcolors{2}{gray!25}{white} 
\resizebox{\linewidth}{!}{
\begin{tabular}{|c|c|c|c|c|c|c|}
\hline
\textbf{Name} & \textbf{Phising Source} & \textbf{Legitimate Source} & \textbf{Collection date} & \textbf{Total samples} & \textbf{Phishing Samples} & \textbf{Legitimate Samples} \\
\hline
Cui et al. \cite{cui2017tracking} & PhishTank  & Alexa Top rank & 2016 & 43806 & 19066 & 24800 \\
PILWD \cite{sanchez2022phishing} &  Phishtank & \makecell{Quantcast Top Sites \\ and The Majestic Million} & 2019-2020 & 132000 & 65613 & 66964 \\
Opara et al. \cite{opara2024look} & \makecell{Korkmaz \\et al. \cite{korkmaz2020deep}} &  Alexa Top rank & 2019 & 45374 & 22687 & 22687 \\
Aljofey et al. \cite{aljofey2022effective} & PhishTank & Stuf Gate42 & 2020 & 60252 & 27280 & 32972 \\
\hline
\end{tabular}}
\caption{Phishing Dataset Information}
\label{tab:phishing-dataset}
\end{table}

\subsection{Metrics}
To evaluate our classification models, we use a set of metrics commonly applied in classification tasks, specifically in phishing detection tasks \cite{karim2023phishing, sahingoz2019machine, sanchez2022phishing, opara2024look}. Specifically, we consider Accuracy, F1-score, Precision, and Recall as they provide valuable understandings of model performance. However, our primary focus will be accuracy and F1-score since these two metrics are widely used in state-of-the-art classification problems when dealing with balanced datasets. 

\subsection{Experiments}

Next, we describe the experiments designed to evaluate our proposal. The first experiment evaluates several classification models, including DT, KNN, XGB, SVM, and RF, using the dataset from Cui et al. \cite{cui2017tracking} The goal is to identify the best-performing model based on the chosen evaluation metrics.

In the second experiment, we compare the proposed method against the baseline model. We conduct a direct comparison using the same dataset to train and test the models, ensuring the approaches are under identical conditions. We reduce the training set size to investigate how the amount of training data affects model performance.

The training set reductions are increments from  $100\%$ to $50\%$, $25\%$, $10\%$, and 5\%. This allows us to assess how each model performs with varying amounts of data, helping to highlight the robustness and adaptability of the approaches when only limited training samples are available. It is important to mention that this reduction applies only to the training and validation sets, while the number of test samples remains unchanged across all experiments. This ensures that the observed performance differences are due solely to changes in the training data size without introducing variability in the test conditions.

The third experiment evaluates the robustness of the method against potential adversarial attacks. Specifically, it tests whether the model can maintain accuracy when exposed to manipulated input data. In this experiment, we introduce an injection attack by placing 2,000 words from a different class at the beginning of each sample; the data injected is selected from a different dataset than the one used to train the model, ensuring that the prediction of the model is not influenced by prior knowledge of the injected data. The purpose of this injection attack is to measure the ability of the models to classify samples even when there is misleading information. As in the previous experiment, we apply the training set reduction. Furthermore, 

The dataset is divided using a 74:16:20 split to conduct the experiments, with $74\%$ for training, $16\%$ for validation, and $20\%$ for testing. Additionally, we extract $20\%$ of the training samples to create a subset for validation. The experiments also use a 5-fold cross-validation methodology to guarantee a robust and reliable evaluation of the models. Each subset of the data is used sequentially in a row for training and testing the model, mitigating problems related to overfitting and bias.

\section{Experimental Results}
\label{sec:results}
The objective of the first experiment is to select the model with the highest performance in phishing attack classification using BoW as input. Table \ref{tab:model_performance} shows the performance of the five selected models in terms of accuracy, F1 score, Precision and recall on the dataset proposed by Cui et al. \cite{cui2017tracking}.

\begin{table}[h!]
\centering
\resizebox{0.8\linewidth}{!}{
\begin{tabular}{|l|c|c|c|c|}
\hline
\textbf{Model} & \textbf{Accuracy} & \textbf{F1-Score} & \textbf{Precision} & \textbf{Recall} \\ \hline
Decision Tree Classifier   & 95.90  & 95.87  & 95.92  & 96.56  \\ \hline
K Neighbors Classifier     & 87.88  & 87.87  & \textbf{99.05}  & 78.41  \\ \hline
XGB                    & 97.09  & 97.07  & 97.80  & 96.79  \\ \hline
SVM                        & 93.32  & 93.27  & 93.48  & 94.27  \\ \hline
\textbf{Random Forest}              & \textbf{97.82}  & \textbf{97.81}  & 98.36  & \textbf{97.62}  \\ \hline
\end{tabular}}
\caption{Performance comparison of the state-of-the-art models. The dataset used for this evaluation is the one proposed by Cui et al. \cite{cui2017tracking}.}
\label{tab:model_performance}
\end{table}

We can see that RF achieves the highest performance with a $97.81\%$ F1-Score and $97.82\%$ accuracy, outperforming the rest of the models. XGB ranks second with a $97.07\%$ F1-score and $97.09\%$ accuracy. In contrast, the KNN algorithm performs poorly in this task, achieving only $87.87\%$ F1-Score and $87.88\%$ accuracy. Based on these results, we integrated RF as the model for BoW classification in the key component extraction pipeline for the following experiments.

The objective of the second experiment is to compare the proposed method against the baseline method. Table  \ref{tab:standard-test} presents the results across different scenarios regarding accuracy and F1-Score. The column "Reduction" shows how the training set and the validation set are reduced, and the "Train" column contains the number of samples of the reduced set. The table includes the standard deviation (STD) values for the F1-score and Accuracy values, obtained by calculating the reported values using 5-fold cross-validation.

\begin{table}[ht]
\centering
\resizebox{\linewidth}{!}{
\begin{tabular}{|c|c|c|c|c|c|c|c|}
\hline
\textbf{Dataset} & \textbf{Reduction} & \textbf{Train} & \textbf{Test} & \multicolumn{2}{c|}{\textbf{PhishKey Extractor}} & \multicolumn{2}{c|}{\textbf{Opara et al. \cite{opara2024look}}} \\
\textbf{} & \textbf{} & \textbf{(Validation)} & \textbf{} & \textbf{ACC (STD)} & \textbf{F1 (STD)} & \textbf{ACC (STD)} & \textbf{F1 (STD)} \\
\hline
Cui et al. \cite{cui2017tracking} & 100 & 28036 (7009) & 8761 & 97.70 (0.18) & 97.69 (0.18) & \textbf{\textcolor{darkgreen}{98.01 (0.04)}} & \textbf{\textcolor{darkgreen}{97.69 (0.04)}} \\
 & 50 & 14018 (3504) &  & 96.97 (0.25) & 96.95 (0.25) & \textbf{\textcolor{darkgreen}{97.17 (0.38)}} & \textbf{\textcolor{darkgreen}{97.15 (0.39)}} \\
 & 25 & 7009 (1752) &  & 96.17 (0.30) & 96.13 (0.30) & \textbf{\textcolor{darkgreen}{96.88 (0.54)}} & \textbf{\textcolor{darkgreen}{96.86 (0.54)}} \\
 & 10 & 2804 (700) &  & 95.28 (0.28) & 95.24 (0.28) & \textbf{\textcolor{darkgreen}{96.20 (0.10)}} & \textbf{\textcolor{darkgreen}{96.18 (0.11)}} \\
 & 5 & 1402 (350) &  & 94.79 (0.28) & 94.75 (0.28) & \textbf{\textcolor{darkgreen}{95.38 (0.48)}} & \textbf{\textcolor{darkgreen}{95.36 (0.48)}} \\
\hline
PILWD \cite{sanchez2022phishing} & 100 & 84480 (21120) & 26400 & \textbf{\textcolor{darkgreen}{97.82 (0.13)}} & \textbf{\textcolor{darkgreen}{97.82 (0.13)}} & 97.11 (0.49) & 97.11 (0.49) \\
 & 50 & 42424 (10607) &  & \textbf{\textcolor{darkgreen}{97.28 (0.15)}} & \textbf{\textcolor{darkgreen}{97.28 (0.15)}} & 96.63 (0.20) & 96.63 (0.20) \\
 & 25 & 21120 (5280) &  & \textbf{\textcolor{darkgreen}{96.46 (0.24)}} & \textbf{\textcolor{darkgreen}{96.46 (0.24)}} & 95.50 (0.61) & 95.50 (0.62) \\
 & 10 & 8448 (2112) &  & \textbf{\textcolor{darkgreen}{95.29 (0.21)}} & \textbf{\textcolor{darkgreen}{95.29 (0.21)}} & 94.34 (0.36) & 94.33 (0.37) \\
 & 5 & 4224 (1056) &  & \textbf{\textcolor{darkgreen}{94.12 (0.34)}} & \textbf{\textcolor{darkgreen}{94.11 (0.34)}} & 93.05 (0.37) & 93.05 (0.37) \\
\hline
Opara et al. \cite{opara2024look} & 100 & 29038 (7260) & 9075 & \textbf{\textcolor{darkgreen}{98.70 (0.16)}} & \textbf{\textcolor{darkgreen}{98.70 (0.16)}} & 98.21 (0.47) & 98.21 (0.47) \\
 & 50 & 11615 (2904) &  & 97.82 (0.34) & 97.82 (0.34) & \textbf{\textcolor{darkgreen}{98.07 (0.33)}} & \textbf{\textcolor{darkgreen}{98.07 (0.33)}} \\
 & 25 & 5807 (1452) &  & 97.00 (0.16) & 97.00 (0.16) & \textbf{\textcolor{darkgreen}{97.43 (0.19)}} & \textbf{\textcolor{darkgreen}{97.43 (0.19)}} \\
 & 10 & 2322 (685) &  & \textbf{\textcolor{darkgreen}{95.99 (0.16)}} & \textbf{\textcolor{darkgreen}{95.99 (0.16)}} & 95.94 (0.90) & 95.94 (0.90) \\
 & 5 & 1160 (291) &  & 94.55 (0.58) & 94.55 (0.58) & \textbf{\textcolor{darkgreen}{94.79 (0.65)}} & \textbf{\textcolor{darkgreen}{94.79 (0.65)}} \\
\hline
Aljofey et al. \cite{aljofey2022effective} & 100 & 38562 (9640) & 12050 & \textbf{\textcolor{darkgreen}{95.37 (0.22)}} & \textbf{\textcolor{darkgreen}{95.27 (0.22)}} & 94.43 (0.85) & 94.33 (0.86) \\
 & 50 & 19281 (4820) &  & \textbf{\textcolor{darkgreen}{93.89 (0.27)}} & \textbf{\textcolor{darkgreen}{93.77 (0.27)}} & 92.02 (0.50) & 91.89 (0.48) \\
 & 25 & 9640 (2410) &  & \textbf{\textcolor{darkgreen}{92.05 (0.45)}} & \textbf{\textcolor{darkgreen}{91.90 (0.45)}} & 90.48 (0.90) & 90.29 (0.86) \\
 & 10 & 3856 (964) &  & \textbf{\textcolor{darkgreen}{89.57 (0.56)}} & \textbf{\textcolor{darkgreen}{89.32 (0.57)}} & 88.47 (1.08) & 88.19 (1.11) \\
 & 5 & 1928 (482) &  & \textbf{\textcolor{darkgreen}{87.90 (0.69)}} & \textbf{\textcolor{darkgreen}{87.54 (0.75)}} & 87.03 (0.56) & 86.74 (0.53) \\
\hline
\end{tabular}}
\caption{Proposed method comparison against the state-of-the-art baseline proposed by Opara et al. \cite{opara2024look}. The best results per row are highlighted in green.}
\label{tab:standard-test}
\end{table}

The results indicate that PhishKey Extractor achieves the highest accuracy and F1 scores across various datasets and their reduction tests, consistently matching or outperforming the baseline model proposed by Opara et al. \cite{opara2024look}. Specifically, PhishKey demonstrated superior performance on datasets presented by Sanchez-Paniagua et al. and Aljofey et al., with improvements ranging from $0.65$ to $1.07$ percentage points on the Sanchez-Paniagua dataset and from $0.87$ to $1.87$ percentage points on the Aljofey dataset \cite{aljofey2022effective}. In contrast, the baseline method outperformed PhishKey on the dataset proposed by Cui et al., with a performance gain between $0.20$ and $0.92$ percentage points.

Furthermore, PhishKey achieves competitive results with lower and more stable standard deviations (STD) across multiple scenarios. Specifically, the STD values of PhishKey range from $0.13\%$ to $0.69\%$, while the baseline STD of the method ranges from $0.04\%$ to $1.08\%$. This range of STD values indicates that PhishKey provides a more stable performance across the 5-fold cross-validation experiments. Lower and more stable STD values reflect the consistency of the model and predictability, reducing the possibilities of performance fluctuations across different datasets.

The purpose of the third experiment is to evaluate the robustness of the methods against potential adversarial attacks. This injection attack measures the ability of the models to classify samples accurately, even with misleading information. We introduce an injection attack by placing 2000 words from a sample of the opposite class of a different dataset at the beginning of each test sample. We apply this experiment to our method PhishKey and the base line method proposed by Opara et al. \cite{opara2024look}. 

Table \ref{tab:injection-test} presents the results across different scenarios regarding Accuracy and F1-Score and presents relevant information about the reduction test and the STD values related to the 5-fold cross-validation.

\begin{table}[ht!]
\centering
\resizebox{\linewidth}{!}{
\begin{tabular}{|c|c|c|c|c|c|c|}
\hline
\textbf{Dataset} & \textbf{Injection Source} & \textbf{Reduction} & \multicolumn{2}{c|}{\textbf{PhishKey}} & \multicolumn{2}{c|}{\textbf{Opara et al. \cite{opara2024look}}} \\
\textbf{} & \textbf{} & \textbf{} & \textbf{ACC (STD)} & \textbf{F1 (STD)} & \textbf{ACC (STD)} & \textbf{F1 (STD)} \\
\hline
Cui et al. \cite{cui2017tracking} & PILWD \cite{sanchez2022phishing} & 100& \textbf{\textcolor{darkgreen}{96.38 (0.29)}} & \textbf{\textcolor{darkgreen}{96.37 (0.29)}} & 84.38 (\textcolor{red}{2.73}) & 84.09 (\textcolor{red}{2.93}) \\
 &  & 50 & \textbf{\textcolor{darkgreen}{95.57 (0.93)}} & \textbf{\textcolor{darkgreen}{95.55 (0.93)}} & 82.74 (\textcolor{red}{2.08}) & 82.06 (\textcolor{red}{2.61}) \\
 &  & 25 & \textbf{\textcolor{darkgreen}{94.52 (0.68)}} & \textbf{\textcolor{darkgreen}{94.49 (0.67)}} & 84.27 (\textcolor{red}{2.15}) & 84.00 (\textcolor{red}{2.23}) \\
 &  & 10 & \textbf{\textcolor{darkgreen}{92.56 (\textcolor{red}{3.40})}} & \textbf{\textcolor{darkgreen}{92.53 (\textcolor{red}{3.39})}} & 80.35 (\textcolor{red}{4.46}) & 79.93 (\textcolor{red}{4.72}) \\
 &  & 5  & \textbf{\textcolor{darkgreen}{92.12 (\textcolor{red}{1.46})}} & \textbf{\textcolor{darkgreen}{92.10 (\textcolor{red}{1.44})}} & 79.65 (\textcolor{red}{4.38}) & 79.18 (\textcolor{red}{4.53}) \\
\hline
PILWD \cite{sanchez2022phishing} & Cui et al. \cite{cui2017tracking} & 100& \textbf{\textcolor{darkgreen}{93.10 (\textcolor{red}{1.65})}} & \textbf{\textcolor{darkgreen}{93.06 (\textcolor{red}{1.68})}} & 69.71 (\textcolor{red}{2.04}) & 69.59 (\textcolor{red}{2.11}) \\
 &  & 50& \textbf{\textcolor{darkgreen}{93.77 (0.40)}} & \textbf{\textcolor{darkgreen}{93.75 (0.40)}} & 68.75 (\textcolor{red}{4.29}) & 68.63 (\textcolor{red}{4.37}) \\
 &  & 25& \textbf{\textcolor{darkgreen}{93.54 (0.38)}} & \textbf{\textcolor{darkgreen}{93.52 (0.39)}} & 69.86 (\textcolor{red}{5.02}) & 69.33 (\textcolor{red}{5.15}) \\
 &  & 10  & \textbf{\textcolor{darkgreen}{92.60 (0.23)}} & \textbf{\textcolor{darkgreen}{92.59 (0.24)}} & 67.62 (\textcolor{red}{2.39}) & 67.28 (\textcolor{red}{2.57}) \\
 &  & 5 & \textbf{\textcolor{darkgreen}{92.18 (0.44)}} & \textbf{\textcolor{darkgreen}{92.18 (0.44)}} & 71.13 (\textcolor{red}{4.31}) & 71.03 (\textcolor{red}{4.28}) \\
\hline
Opara et al. \cite{opara2024look} & Aljofey et al. \cite{aljofey2022effective} & 100 & \textbf{\textcolor{darkgreen}{97.47 (0.33)}} & \textbf{\textcolor{darkgreen}{97.47 (0.33)}} & 87.87 (\textcolor{red}{2.36}) & 87.78 (\textcolor{red}{2.50}) \\
 &  & 50  & \textbf{\textcolor{darkgreen}{95.57 (\textcolor{red}{1.50})}} & \textbf{\textcolor{darkgreen}{95.56 (\textcolor{red}{1.52})}} & 86.90 (0.65) & 86.82 (0.66) \\
 &  & 25& \textbf{\textcolor{darkgreen}{95.59 (0.90)}} & \textbf{\textcolor{darkgreen}{95.58 (0.91)}} & 83.51 (\textcolor{red}{2.60}) & 83.49 (\textcolor{red}{2.61}) \\
 &  & 10  & \textbf{\textcolor{darkgreen}{94.31 (0.34)}} & \textbf{\textcolor{darkgreen}{94.31 (0.34)}} & 72.02 (\textcolor{red}{7.76}) & 71.41 (\textcolor{red}{7.75}) \\
 &  & 5  & \textbf{\textcolor{darkgreen}{93.07 (0.77)}} & \textbf{\textcolor{darkgreen}{93.06 (0.77)}} & 70.11 (\textcolor{red}{3.33}) & 69.27 (\textcolor{red}{3.54}) \\
\hline
Aljofey et al. \cite{aljofey2022effective} & Opara et al. \cite{opara2024look} & 100  & \textbf{\textcolor{darkgreen}{94.36 (0.22)}} & \textbf{\textcolor{darkgreen}{94.22 (0.22)}} & 86.93 (\textcolor{red}{3.23}) & 86.71 (\textcolor{red}{3.19}) \\
 &  & 50 & \textbf{\textcolor{darkgreen}{92.53 (0.47)}} & \textbf{\textcolor{darkgreen}{92.34 (0.46)}} & 81.62 (\textcolor{red}{2.03}) & 81.33 (1.98) \\
 &  & 25 & \textbf{\textcolor{darkgreen}{90.61 (\textcolor{red}{1.10})}} & \textbf{\textcolor{darkgreen}{90.38 (\textcolor{red}{1.17})}} & 81.37 (1.79) & 80.93 (1.57) \\
 &  & 10 & \textbf{\textcolor{darkgreen}{86.83 (\textcolor{red}{2.07})}} & \textbf{\textcolor{darkgreen}{86.35 (\textcolor{red}{2.23})}} & 80.48 (2.03) & 79.91 (1.87) \\
 &  & 5& \textbf{\textcolor{darkgreen}{86.45 (0.69)}} & \textbf{\textcolor{darkgreen}{85.96 (0.70)}} & 77.67 (\textcolor{red}{4.01}) & 77.10 (\textcolor{red}{3.95}) \\
\hline
\end{tabular}}
\caption{Injection Test Results against the state-of-the-art baseline proposed by Opara et al. \cite{opara2024look}. The green color highlights the best results per row and the red highlights the highest standard deviation (STD) values. The "Dataset" column shows the data source to evaluate, and the "Injection" column indicates the source of the injected samples}
\label{tab:injection-test}
\end{table}

Across the datasets and their reduction tests, PhishKey outperforms the baseline method in both accuracy and F1 score. The most significant performance gap appears in the PILWD and Aljofey et al. datasets, where our method achieves an accuracy of up to $96.38\%$ with a low STD of $0.29\%$. In contrast, the baseline reaches only $84.38\%$ accuracy, with a much higher STD of $2.73\%$.
In terms of stability, PhishKey  shows a lower STD compared to the baseline. In a case where an attacker manipulates the initial part of a sample before its submission to the model, our approach shows the slightest reduction in performance with drops ranging from $1.32$ and $8.53$ percentage points, as compared to the results of the second experiment demonstrated in Table \ref{tab:standard-test}. In contrast, the base performance of the model drops significantly, with reductions ranging from $8.80$ to $27.88$ percentage points.

The results of this experiment indicate consistent performance across folds in cross-validation, suggesting that PhishKey  is more robust and less affected by dataset variation than the baseline model. These findings support the conclusion that our method is more robust against sample manipulation and injection attacks, as it uses the whole input and learns more effectively to differentiate the essential parts of each sample during classification.

\section{Conclusions}
\label{sec:Conclusions}
Phishing detection remains a complex challenge due to the constant evolution of phishing techniques, which require adaptable and robust classification methods. 

Across the multiple approaches, automatic feature extraction methods have an inherent advantage over other algorithms due to their capability to adapt to the constant evolution of phishing attacks.

This work introduces PhishKey, a novel approach that automatically extracts features from URL and HTML sources in phishing webpages, making the detection process more adaptable and effective.

PhishKey employs a hybrid structure: it uses character-level processing with CNN and fully connected layers (FCL) for URLs. At the same time, we propose CAPE (Centroid-Based Key Component Phishing Extractor) for the HTML processing. CAPE analyzes HTML sources at the word level, starting with a tokenization process and converting each word in the document into 100-dimensional vectors. After selecting the $2000$ most relevant words, a model performs the classification using them as a BoW input. The final output is generated by combining the URL and HTML classifications through a soft-voting ensemble method.

The experimental results show that PhishKey achieves state-of-the-art results, with accuracy and F1-score reaching $98.70\%$,  with an increase of $0.5$  percentage points over the baseline method. In addition, PhishKey shows lower STD values across folds, which indicates consistent and stable performance, which is critical for phishing detection, where variability can compromise reliability.

PhishKey presents several  important advantages in phishing detection:

\textbf{Automatic feature extraction}. As mentioned, PhishKey employs automated feature extraction, allowing it to adapt to evolving phishing strategies without extensive manual intervention. It will not require major modifications, such as new features when new samples are released,  as a manual approach would require.

\textbf{Full document utilization}. 
Unlike other methods that often truncate HTML content to meet model input requirements, PhishKey ensures the entire document is processed, maintaining relevant information and preventing data loss during preprocessing. This approach minimizes information loss and enhances the adaptability of the model to evolving attack strategies.

\textbf{Robustness against manipulated samples or data injection attacks}.
PhishKey demonstrates robustness to adversarial manipulations, such as injection attacks, by automatically identifying and prioritizing key features in each sample. Experiment 3 in Section \ref{sec:results},  confirms the superior performance of PhishKey under adversarial conditions, with minimal performance degradation compared to the baseline.

\textbf{Consistent performance with stable and low STD values}. PhishKey achieves lower and more stable STD values, reflecting its reliability and consistency across different datasets. This stability is critical to handle diverse and dynamic real-world scenarios for phishing detection.

Dynamic adaptation to threats without manual modifications through automatic feature extraction, full document processing to preserve contextual integrity, and resilience to adversarial manipulation are characteristics needed to address phishing classification. These characteristics are particularly critical in the constantly evolving phishing landscape, where attackers are developing new strategies and techniques to evade detection models every day.

In future work, we aim to extend our research in four key directions: First, we plan to address the challenges of extracting key components from HTML by pre-training a transformer-based model on a large HTML dataset using advanced natural language techniques. Second, we will optimize our method by developing lightweight models for word embeddings and classifiers, enabling real-time phishing detection systems on resource-constrained devices. Third, we will study the interpretability of the model to understand the decision-making process behind phishing classification. Finally, we aim to explore the classification and attribution of phishing attacks via phishing kits as we previously explore \cite{castano2023phikita}, offering insights into attack authorship and contributing to threat intelligence and prevention efforts.

\section{Acknowledgements}
This work has been partially funded by the Recovery, Transformation, and Resilience Plan, financed by the European Union (Next Generation) thanks to the LUCIA project (Fight against Cybercrime by applying Artificial Intelligence) granted by INCIBE to the University of León.

This work has been partially funded by the Horizon Program of the European Union under the project KINAITICS (AI for cybersecurity reinforcement), Grant Agreement No. 101070176)

\bibliographystyle{cas-model2-names}

\bibliography{references}

\end{document}